\pdfoutput=1 

\documentclass[aps,prl,superscriptaddress,12pt, longbibliography,reprint]{revtex4-2}

\makeatletter
\renewcommand{\p@subsection}{}
\renewcommand{\p@subsubsection}{}
\makeatother

\usepackage{amsmath,amsfonts,amssymb,amsthm}
\usepackage{mathtools}
\usepackage{bm}
\usepackage{setspace}
\usepackage[normalem]{ulem} 
\usepackage{stmaryrd}       
\usepackage{dsfont}         
\usepackage{expdlist}
\usepackage{physics}
\usepackage{verbatim}
\usepackage{xspace}
\usepackage{subfigure}

\usepackage[page]{appendix} 
\usepackage{titlesec}
\titleformat*{\section}{\large\bfseries\center}

\newcommand{\RNum}[1]{\uppercase\expandafter{\romannumeral #1\relax}}

\usepackage{breakurl}

\usepackage{graphicx,xcolor}
\usepackage{hyperref}
\hypersetup{pdfstartview=}
\usepackage{url}

\usepackage{verbatim}
\usepackage{alltt}
\usepackage{moreverb}

\usepackage{xr}

\providecommand{\ignore}[1]{}

\newif\ifcmnt
\cmnttrue
\ifdefined\cmntsoff\cmntfalse\fi

\ifcmnt

\else

\fi

\newcommand{\defeq}{\doteq}

\newcommand{\cN}{\mathcal{N}}

\newtheorem*{theorem*}{Theorem}

\newtheorem*{problem*}{Problem}

\newtheorem*{lemma*}{Lemma}

\newcommand{\Coeff}{\textrm{Coeff}}

\begin{document}

\title{Multi-mode Gaussian State Analysis with one Bounded Photon Counter}

\author{Arik Avagyan}
\affiliation{National Institute of Standards and Technology, Boulder, Colorado 80305, USA}
\affiliation{Department of Physics, University of Colorado, Boulder, Colorado, 80309, USA}
\author{Emanuel Knill}
\affiliation{National Institute of Standards and Technology, Boulder, Colorado 80305, USA}
\affiliation{Center for Theory of Quantum Matter, University of Colorado, Boulder, Colorado 80309, USA}
\author{Scott Glancy}
\affiliation{National Institute of Standards and Technology, Boulder, Colorado 80305, USA}

\begin{abstract} 

Gaussian states are ubiquitous in quantum optics and information
processing, and it is essential to have effective tools for their
characterization. One such tool is a photon-number-resolving
detector, and the simplest configuration involves counting the total
number of photons in the state to be characterized. This motivates the
following question: What properties of a multi-mode Gaussian state are
determined by the signal from one detector that measures total number
photons up to some bound? We find that if the Gaussian state occupies
$S$ modes and the probabilities of $n$ photons for all $n\leq 8S$ are
known, then we can determine the spectrum of the Gaussian covariance
matrix and the magnitude of the displacements in each eigenspace of the covariance
matrix. Nothing more can be learned, even if all photon-number
probabilities are known. When the state is pure, the covariance
matrix spectrum determines the squeezing parameters of the state.
\end{abstract} 

\maketitle



Gaussian states play important roles in most applications of quantum
optics. Besides occurring naturally in the form of displaced vacuum
and thermal states, Gaussian states, particularly squeezed ones, are
important resources in quantum communication, quantum sensing and
all-optical quantum computing~\cite{wang2007quantum,
  weedbrook:qc2012a}. To support these applications, it is necessary
to characterize Gaussian states in experiments. Common tools for such
characterizations include homodyne and heterodyne measurements
\cite{weedbrook:qc2012a, paris2003purity,laurat2005entanglement,
  dauria2005characterization,porzio2007homodyne, rehacek2009effective,
  dauria2009full, paternostro:qc2009a, buono2010quantum,
  blandino:qc2012a, esposito2014pulsed}. These measurements obtain
information about the quadratures of the optical state and can be used
to fully characterize any state. Instead of measuring the quadratures,
one can use photo-detectors or photon-number-resolving (PNR) detectors
in combination with linear elements such as beam splitters and phase
shifters. There are many proposals and demonstrations for characterizing
Gaussian states with various configurations of these elements ~\cite{fiurasek2004how,wenger2004pulsed,wallentowitz1996unbalanced,konrad1996direct,manko2003photon,manko2009photon,leibfried1996exp,nogues2000meas},
in some cases demonstrating sensitivity advantages of PNR detectors
for measuring specific characteristics of a state such as its
phase~\cite{wu2019quantum}, displacement~\cite{burd2019quantum}, 
squeezing~\cite{burenkov2017full,zuo2022determination}, or
temperature~\cite{bezerra2021quadrature}.  Full
characterization of a Gaussian multi-mode state is possible with PNR
detectors by means of measurements involving multiple configurations
of linear elements~\cite{parthasarathy2015from,kumar2020optimal}.

The simplest measurement setup with a PNR detector involves directly
measuring the total photon number of a multi-mode Gaussian state. In the rest of the manuscript we omit
the adjective ``total" when referring to the total photon-number
observable or to the total photon-number distribution of a Gaussian
state. Given the covariance matrix and the displacement for the
Gaussian state, it is possible to compute the photon-number
probability distribution. Expressions for the probability of a given
number of photons have been obtained in
Ref.~\cite{dodonov1994multi}. We consider the question of what can be
learned about the multi-mode Gaussian state from the probabilities of
$0,\ldots, N$ photons. In Ref.~\cite{qc:avagyan2022} we showed that
the photon-number distribution is determined by the spectrum of the
state's covariance matrix and the absolute displacements in each
eigenspace. We used a non-constructive analysis to show that the
infinitely many exact photon-number probabilities determine the
spectrum and absolute displacements. Here we show that there is an
efficient way to determine the spectrum and absolute displacements
from finitely many photon-number probabilities. If the number of modes
of the Gaussian state is $S$, $4S$ parameters suffice to describe the
covariance matrix spectrum and absolute displacement. We prove that
the probabilities of $k$ photons for $k=0,\ldots,8S$ are
sufficient. Given these $8S+1$ exact probabilities, we show how to
efficiently determine the covariance matrix spectrum and the absolute
displacements by algebra and root-finding. If the Gaussian state is
pure, the squeezing parameters can also be deduced.  Thus, simple
photon counting with counters able to statistically resolve only a
bounded number of photons suffices to determine useful diagnostics of
Gaussian states. Further work is required to determine statistically
robust and effective methods for learning Gaussian state parameters
from estimated photon-number probabilities.

\noindent \emph{Preliminaries.} We consider Gaussian states of $S$
modes defined by $S$ mode annihilation operators
$\hat a_1,\ldots, \hat a_S$. Each mode has two standard quadrature
operators. For mode $k=1,\ldots, S$, we write
$\hat r_{2k-1} = (\hat a_k+\hat a^\dagger _k)/\sqrt{2}$ and
$\hat r_{2k} = (\hat a_k - \hat a^\dagger_k)/(i\sqrt{2})$ for its
``$q$'' and ``$p$'' quadratures. The $S$ modes are measured by a
photon counter. The expectation of an operator $A$ for state $\rho$ is
written as $\langle A\rangle_{\rho}$. If the state $\rho$ is clear in
context, we omit it from the notation. An ideal photon counter returns
the number $n$ if exactly $n$ photons occupy the $S$ modes.  For a
general state $\rho$, the probability that it returns $n$ is
$p_n \defeq \langle \Pi_n\rangle = \tr(\rho \Pi_n)$, where $\Pi_n$ is
the projector onto the subspace of states with $n$ photons in
total. We require that the needed values of $p_{n}$ can be inferred
from the available detector's outcome distribution. The detector may
be an inefficient or approximate photon counter.  States of $S$ modes
can be described by their Wigner function.  The Wigner function
$W(\vec{r})$ is a function on the $2S$-dimensional phase space of $S$
modes. Here, the vector $\vec{r} = (r_1,\ldots, r_{2S})$ denotes a
  point in phase space, where for $k=1,\ldots S$, $r_{2k-1}$ and
$r_{2k}$ are the two phase space coordinates for mode $k$.  We
  write \(r= |\vec{r}|\) for the length of the vector \(\vec{r}\).
Gaussian states have Gaussian Wigner functions. A Gaussian state
$\rho$ of $2S$ modes is determined by its displacement $\vec{d}$
  and its optical covariance matrix $\Gamma$.
The entries of \(\vec{d}\) are $d_k = \langle \hat r_k\rangle$.
We define the covariance matrix according to
$\Gamma_{kl} = \langle\hat r_k \hat r_l + \hat r_l \hat r_k\rangle -
2\langle\hat r_k\rangle\langle\hat r_l\rangle$.
We write $\rho(\Gamma, \vec{d})$ for the
Gaussian state with covariance matrix $\Gamma$ and displacement
$\vec{d}$.  The Wigner function of $\rho(\Gamma,\vec{d})$ is given by
\begin{align}
    W(\vec{r}) &= \frac{1}{\pi^S} \frac{1}{\sqrt{\det(\Gamma)}} e^{-(\vec{r}-\vec{d})^T \Gamma^{-1}  (\vec{r}-\vec{d})}. \label{eq:wigner}
\end{align}
(See Eq. (20), Sect. 2 A~\cite{weedbrook:qc2012a}, the expression is
adjusted for our conventions.) The Wigner function is a Gaussian
  probability distribution. Our definition of the optical covariance
  matrix implies that this probability distribution's covariance matrix
is \(\Gamma/2\).
For one mode, the Wigner function of
the state $\ket{n}$ with exactly $n$ photons is given by
\cite{leonhardt:qc1997a}:
\begin{align}
  W^{(1)}_n(\vec{r}) &= \frac{(-1)^n}{\pi}e^{-r^2}L_n(2r^2),\label{eq:nwigner}
\end{align}
where $L_n$ is the $n$'th Laguerre polynomial. The
probability of $n$ photons in one mode is given by $ 2\pi \int_{\vec r} d^2\vec{r}\;W^{(1)}_n(\vec{r}) W(\vec{r})$. For our
analysis we need the generating function for Laguerre polynomials
\begin{align}
  L(r;z)&=\sum_{n=0}^{\infty} z^{n}L_{n}(r) = e^{-z
      r/(1-z)}/(1-z).\label{eq:glaguerre}
\end{align}
We denote the degree \(n\) coefficient of a generating function or polynomial \(F(z)\)
by \(\Coeff_{n,z}(F(z))\).

\noindent\emph{Main Result.}  The photon-number distribution of $\rho(\Gamma,\vec{d})$ only
depends on the spectrum of $\Gamma$ and the absolute values of the
displacements within the eigenspaces of $\Gamma$ (see below). We
define the ``normal parameters" of $\Gamma,\vec{d}$ to consist of i)
the distinct eigenvalues of $\Gamma$ in decreasing order
$\bm{\lambda}=(\lambda_k)_{k=1}^h$, where $1\leq h\leq 2S$, ii) the
associated multiplicities $\bm{m}=(m_k)_{k=1}^h$ with
$\sum_{k=1}^{h} m_k = 2S$, and iii) the absolute displacements in the
corresponding eigenspaces $\bm{c}=(c_k)_{k=1}^h$ with $c_k\geq 0$. The
normal parameters can be determined from $\Gamma$ and $\vec{d}$ by
diagonalizing $\Gamma$ with an orthogonal transformation so that
$\Gamma$ has the form $\mathrm{diag}(\gamma_1, ..., \gamma_{2S})$ with
the $\gamma_k$ in decreasing order. The distinct $\gamma_k$ form the
normal parameter $\bm{\lambda}$, and the multiplicities make up
$\bm{m}$. The transformation can be chosen so that the corresponding
transformation of $\vec{d}$ has exactly one entry in each block of
identical $\gamma_k$ which is non-negative, while the other entries
are zero. That entry is the absolute displacement for the block's
diagonal entry and contributes the corresponding entry in
$\bm{c}$. Our main result can be stated as follows:

\noindent\textbf{Theorem:} For Gaussian states, the photon-number
probability distribution is a function of the normal parameters, and
the normal parameters are a function of the first $8S+1$ photon-number
probabilities $(p_0, \ldots , p_{8S})$.

The first part of the theorem is a consequence of the fact that the
photon-number probabilities are determined by moment-like quantities of the
Wigner function. An argument based on the Husimi distribution is in
Ref.~\cite{qc:avagyan2022}. This reference also shows
non-constructively that the normal parameters are a function of the
full photon-number probability distribution. In the generic case, the
normal parameters have $2S$ distinct eigenvalues and $2S$ non-negative
displacements, which can be varied independently. At least $4S$
photon-number probabilities are therefore required to determine the
normal parameters. We leave open the question of whether the first
$4S$ photon-number probabilities are sufficient.

To prove the theorem, we first establish a linear relationship between
the first $k$ photon-number probabilities and the first $k$ moments of
$r^2$ of a probability distribution related to the
Wigner function. Next, we exploit the fact that the first $k-1$
cumulants are functions of the first $k$ such moments. This reduces
the problem to that of showing that the first $8S$ cumulants determine
the normal parameters. We provide an explicit method to compute the
normal parameters from the first $8S$ cumulants. The steps are
explained in the next paragraphs. Technical details are provided in
the Supplemental Material (SM).

For the phase space variable $\vec{r}$, define
$\vec{x_k}= (r_{2k-1}, r_{2k})$ and $x_k^2 = r_{2k-1}^2+r_{2k}^2$, so
that $\vec{x}_k$ is $k$'th mode's phase-space variable. To establish
the relationship between the moments  and
the photon-number probabilities, let $W_n$ be the Wigner function of
$\Pi_n$, so that
$p_n =  (2\pi)^S\int dr^{2S} W_n(\vec{r})
W(\vec{r})$. 
An explicit form of \(W_{n}(\vec{r})\) is given in SM Eq.~\eqref{wn_explicit}.
We analyze $p_n$ by means of the generating
function $M(\vec{r};z) = \sum_n W_n(\vec{r}) z^n$. For one mode, this generating function can be expressed in terms of the generating
function for Laguerre polynomials as
$M^{(1)}(\vec{r};z) = \frac{1}{\pi} e^{-r^2}L(2r^2;-z)
=\frac{1}{\pi}e^{-r^2}e^{z 2r^2/(1+z)}/(1+z)$. For
$S$ modes, it can be seen that
$M(\vec{r};z) = \prod_{k=1}^S M^{(1)}(\vec{x_k};z)$ (SM Eq.~\eqref{laguerre_gen}), consequently
\begin{align}
  M(\vec{r};z) &= \frac{e^{-r^2}}{\pi^S}\frac{1}{(1+z)^S}e^{2 z r^2/(1+z)}.
  \label{ngenfn}
\end{align}
The coefficient of $z^n$ in $M(\vec{r};z)$ is of the form
$\Coeff_{n,z}(M(\vec{r};z)) = e^{-r^2} P_n(r^2)$, where $P_n(x)$ is a polynomial of degree $n$ with
the highest degree term given by $(2x)^n/(\pi^Sn!) $. In particular,
\(P_{0}(x) = 1/\pi^{S}\).  See SM Eq.~\eqref{pn_explicit} for an
  efficiently computable expression for \(P_{n}(x).\) We have
$W_n(\vec{r}) = \left(\frac{d}{dz}\right)^{n} \frac{M(\vec{r};z)}{n!}
|_{z=0} = e^{-r^2} P_n(r^2)$.  As a result, the probability of $n$
photons is given by
\begin{align}
  p_n &= (2\pi)^S\int d^{2S}\vec{r} e^{-r^2} W(\vec{r}) P_n(r^2).
  \label{eq:momentston}
\end{align}
The integral in this identity is invariant under orthogonal rotations
of phase space. In Eq.~\eqref{eq:wigner} an orthogonal rotation $O$
transforms $\Gamma$ to $O\Gamma O^T$ and $\vec d$ to $O\vec d$.
As described earlier, the transformation can be chosen such that $\Gamma$ is $\Gamma=\mathrm{diag}(\gamma_1,\ldots, \gamma_{2S})$ with the $\gamma_k$ in decreasing order. We can also assume that for a block of
identical $\gamma_k$, the corresponding $d_k$ are zero except for the
lowest index \(k\) in the block where \(d_{k}\) is non-negative. This confirms that the photon-number distribution only depends on the normal
parameters~\cite{qc:avagyan2022}. We now assume that $p_0, \ldots, p_{8S}$ are given and aim to infer the normal parameters.

From Eq.~\eqref{eq:momentston} we obtain
\(p_{0}=2^{S}\int d^{2S}\vec{r} e^{-r^{2}}W(\vec{r})\).  It follows
that \(W'(\vec{r}) = 2^{S}e^{-r^{2}}W(\vec{r})/p_{0}\) is a
probability distribution. From
\(p_{n}/p_{0}=\int d^{2S}\vec{r}W'(\vec{r}) \pi^{S}P_{n}(r^{2})\), we
see that \(p_{n}/p_{0}\) is a linear combination of the first \(n\)
moments of \(r^{2}\) with respect to \(W'(\vec{r})\). Because
\(\Coeff_{n,r^{2}}\left(P_{n}(r^{2})\right) \ne 0\), the expression of
$p_{n}/p_{0}$ in terms of the moments can be inverted to express the
$n$'th moment of $r^2$ as a linear combination of the relative
photon-number probabilities $p_1/p_{0},\ldots,p_n/p_{0}$.

We consider the exponential generating function for the moments of $r^2$,
\begin{align}\label{eq:F(z)}
  F(z) &= \int d^{2S}\vec{r} e^{r^2 z}  W'(\vec{r}).
\end{align}
The coefficient \(\Coeff_{n,z}F(z)\) is $1/n!$ times the $n$'th moment
of $r^2$   Next we substitute the
Gaussian form from Eq.~\eqref{eq:wigner} into the expression for
$F(z)$, which results in a Gaussian integral with exponent
$X(z)= - r^2(1-z) - (\vec r - \vec d)^T \Gamma^{-1} (\vec r-\vec d)$
(see SM Eq.~\eqref{gaussianform}). The integral converges and
is analytic in $z$ for $\Re{z}-1- \gamma_{1}^{-1}<0$. The integral is
\(1\) at $z=0$. Our goal is to determine the logarithmic derivative
$l(z) =(dF(z)/dz)/F(z)$ as a power series in $z$ in a neighborhood of
$z=0$. Let \(f_{n}=\Coeff_{n-1,z}l(z)\). Then \((n-1)! f_{n}\) is
  the \(n\)'th cumulant of the distribution of $r^{2}$.   The first $n$ cumulants
are polynomials of the first $n$ moments, and, therefore, of the first
$n$ relative photon-number probabilities. An efficient recursive
method for obtaining the cumulants is given in
Ref.~\cite{smith1995recursive} and in SM Sect.~\ref{cumulant_comp}. The Gaussian integral in
Eq.~\eqref{eq:F(z)} is a product of $2S$ integrals of the form
$J(z) =C \int dx e^{-x^2(1-z) - (x-d)^2/\gamma}$, where the
  coefficient \(C\) does not depend on \(z\).   The corresponding contribution to
$l(z)$ can be obtained by differentiating under the integral to
express $(dJ/dz) = C \int dx x^{2}e^{-x^2(1-z) - (x-d)^2/\gamma}$,
  which can be evaluated as a second moment for a standard Gaussian
  after a change of variables. 
We obtain
\begin{align}
  \frac{d J(z)/dz}{J(z)} &= \frac{1}{2 (1-z+1/\gamma)} + \left(\frac{d}{\gamma (1-z +1/\gamma)}\right)^2.
\end{align}
Define $\lambda'_k = (1+1/\lambda_k)^{-1}$ and $c'_k = (c_k/\lambda_k)^2$. Then as shown in SM Eq.~\eqref{eq:lexpansion_sm},
\begin{align}
  l(z) &=
  \sum_{k=1}^h\left( \frac{\lambda'_k m_k}{2(1 - \lambda'_kz)} + \frac{\lambda'_k{}^2c'_k}{(1 - \lambda'_kz)^2}\right).
  \label{eq:lexpansion}
\end{align}
The coefficients \(f_{n}\) can now be expressed
  as
\begin{align}
  f_n &=\sum_{k=1}^h \left(\frac{m_k}{2} \lambda'_{k}{}^{n } + n c'_k  \lambda'_k{}^{n+1}\right).\label{fn}
\end{align}
The goal now is to infer the normal parameters given
$\vec{f} = (f_1 ,\ldots, f_{8S})$. The strategy is to first determine
a minimal polynomial \(q(z)\) with \(q(0)=1\) such that
\(\Coeff_{k,z}(q(z)l(z))=0\) for \(4S\leq k \leq 8S-1\).  We show that
\(q(z)l(z)\) is in fact a polynomial of degree at most \(4S-1\) and
that the linear equations that need to be solved are determined by
\(\vec{f}\).  The normal parameters can then be determined by finding
the roots of \(q(z)\) to obtain the \(\lambda'_{k}\) and then
determining \(m_{k}\) and \(c'_{k}\) by solving additional linear
equations.  Consider a polynomial $q(z)$ such that $q(z)l(z)$ is a
polynomial. Inspection of Eq.~\eqref{eq:lexpansion} shows that
$l(z) q(z)$ is a polynomial if and only if $q(z)$ is divisible by
$(1-\lambda'_k z)$ for every $k$ such that $c'_k=0$ and divisible by
$(1-\lambda'_k z)^2$ otherwise.  Define $\cN = \{k: c_k \ne 0\}$ and
let $\bar{h}=|\cN|$.  Thus, the polynomial
$q_0(z)=\prod_{k=1}^{h}(1-\lambda'_k z)\prod_{l\in\cN}(1-\lambda'_lz)$
is the minimum degree polynomial such that $q_0(z)l(z)$ is a
polynomial, \(q_{0}(0)=1\), and every polynomial $q(z)$ such that
$q(z)l(z)$ is a polynomial is divisible by $q_0(z)$.  The degree of
$q_0(z)$ is $h+\bar{h} \leq 2 h$, and the degree of $q_0(z)l(z)$ is at
most $2h-1 \leq 4S-1$.  Suppose that $q(z)$ is a polynomial of degree
at most $4S$ satisfying the condition that
\(\Coeff_{k,z}(q(z)l(z))=0\) for $k=4S,\ldots, 8S-1$. We show that
$q(z)l(z)$ is a polynomial, so $q_0(z)$ divides $q(z)$.  Furthermore,
\(q(z)l(z)\) has degree at most \(4S-1\).  Suppose to the contrary,
and let \(r\) be the minimum power that is greater than or equal to
\(4S\) for which \(\Coeff_{r,z}(q(z)l(z)) \ne 0\).  By assumption
\(r\geq 8S\).  Because the constant term of $q_0(z)$ is $1$ and
$q_0(z)$ has degree at most $4S$, the coefficient
\(\Coeff_{r,z}(q_{0}(z)q(z)l(z)) \ne 0\), which implies that
\(q_{0}(z) q(z) l(z)\) has degree at least \(8S\). But since
\(q_{0}(z) l(z)\) has degree at most \(4S-1\), the degree of
\(q_{0}(z) q(z) l(z)\) is actually at most \(8S-1\).  We conclude that
there can be no \(r\geq 4S\) for which
\(\Coeff_{r,z}(q(z)l(z)) \ne 0\), so \(q(z)l(z)\) is a polynomial of
degree at most \(4S-1\) as claimed.

To determine \(q_{0}(z)\) from \(\vec{f}\), define
$f(z) =\sum_{k=0}^{8S-1} f_{k+1} z^k$, which is the polynomial
obtained by truncating the series for \(l(z)\) at degree \(8S-1\).
Therefore, for any polynomial \(q(z)\) of degree at most \(4S\)
and \(r\leq 8S-1\), \(\Coeff_{r,z}(q(z)l(z))=\Coeff_{r,z}(q(z)f(z))\). It follows
from above that if \(q(z)\) has degree at most \(4S\) and
\(\Coeff_{r,z}(q(z)f(z))=0\) for \(4S\leq r\leq 8S-1\), then
\(q(z)l(z)\) is a polynomial and \(q_{0}(z)\) divides \(q(z)\).
Consequently, to determine \(q_{0}(z)\) from \(\vec{f}\), it sufficies
to find the minimum degree \(q(z)\) such that \(q(0)=1\) and
\(\Coeff_{r,z}(q(z)f(z))=0\) for \(4S\leq r\leq 8S-1\).  This
requires solving \(4S+1\) linear equations in the coefficients of
\(q(z)\).

Once \(q_{0}(z)\) is found, we solve for the roots of \(q_{0}(z)\)
and determine their multiplicities. According to the above, the
roots are the \(1/\lambda_{k}'\) and their multiplicities are either
one or two. The ones with multiplicity two are associated with
nonzero \(c_{k}'\). To solve for \(m_{k}\) and \(c_{k}'\), for
\(1\leq k\leq h\) let \(\vec{u_{k}}\) and \(\vec{v_{k}}\) be the
dimension-\(4S\) vectors whose \(l\)'th entries are
\(\lambda_{k}'{}^{l}\) and \(l\lambda_{k}'{}^{l-1}\), for
\(1\leq l\leq 4S\).  Let $\vec{f'} = (f_{1},\hdots,f_{4S} )$. Then
$\vec{f'} = \sum_k (m_k/2)\vec{u_k} + \sum_{k\in\cN} c_k'
\lambda_k'{}^2\vec{v_k}$. It follows from the theory of Hermite
interpolation that the vectors $\vec u_k$ and $\vec v_k$ are
linearly independent, so there is a unique solution for the
coefficients \(m_{k}\) and \(c_{k}'\) in this identity for
\(\vec{f'}\). The normal parameters can now be determined by
inverting the relationship between the \(\lambda_{k}', c_{k}'\) and
the \(\lambda_{k},c_{k}\).  This completes the proof of the theorem.

\noindent\emph{Calculating the normal parameters.} The proof of the theorem
is constructive. Given the first $8S+1$ photon-number probabilities,
the normal parameters can be computed algebraically. It requires a polynomial
transformation of the photon-number probabilities, root finding and solving
a linear equation. Each step can be efficiently numerically implemented.
Efficient formulas for the necessary transformations are in the SM.

Our results assume exact photon-number probabilities. In an
experimental context, the photon-number probabilities are
estimated. These estimates do not require perfect photon counters.
They can be obtained from inefficient or otherwise noisy
photon detection systems or multiplexed ``click'' photon detectors by
statistical inference after characterizing the detection system. For
a perfectly stable experiment and in the large data limit, the
computed estimates converge to the true probabilities.

The numerical calculation of the normal parameters is highly
non-linear, which may result in numerical instabilities with respect
to the noise in estimated count probabilities.  Instead of direct
numerical calculation, one can use a maximum likelihood method to fit
the estimated photon-number probabilities with the exact ones
calculated from candidate normal parameters. Our theorem implies that
one can use the first \(8S+1\) photon-number
probabilities for the fitting procedure, and the results converge to
the true parameters in the large data limit. Of course, when using
maximum likelihood, all information about the photon number
probabilities should be used to improve the convergence.

\noindent\emph{Interpretation of the normal parameters.}

In the absence of a preferred mode basis, non-displaced Gaussian
states are usually described in terms of their squeezing and thermal
parameters. The covariance matrix $\Gamma$ of a Gaussian state can be
written in the form $O_1^T S O_2^T T O_2 S O_1$, where $O_i$ are
orthogonal matrices associated with passive linear optical
transformations, $T$ is the diagonal covariance matrix of a state that
is thermal in each mode, and $S$ squeezes each mode separately (for
example, see~\cite{serafini2017quantum}).   $T$ is block
diagonal with $2\times 2$ blocks of the form $\tau_k I$, where $I$ is
the $2\times 2$ identity matrix, while $S$ is block diagonal with
$2\times 2$ blocks of the form $e^{\xi_i\sigma_z}$ with
$\sigma_z=\mathrm{diag}(1,-1)$. The thermal parameters are the
$\tau_k \geq 1$ and the squeezing parameters are the $\xi_k\geq 0$.
Except for their ordering, they are determined by the state. In
contrast, the normal parameters give the spectrum of $\Gamma$. This
spectrum constrains the squeezing and thermal parameters but in
general does not determine them. However, if the Gaussian state is
pure, in which case $T$ is the identity matrix, the normal parameters
witness this purity and the squeezing parameters can be deduced from
the spectrum, see Ref.~\cite{qc:avagyan2022}. In general, given the
spectrum, there is a Gaussian state whose covariance matrix has this
spectrum and whose squeezing and thermal parameters can be derived in
a canonical way from the spectrum: If $\gamma_1,\ldots,\gamma_{2S}$ is
the spectrum in non-increasing order, the canonical thermal parameters
are $\sqrt{\gamma_k\gamma_{2S-k-1}}$ for $i=k,\ldots, S$, and the
corresponding squeezing parameters are
$\log(\gamma_i/\gamma_{2S-k+1})/2$. The thermal and squeezing
parameters can vary for different covariance matrices of Gaussian
states with the same canonical thermal and squeezing parameters.

\noindent\emph{Conclusion.}

Squeezed Gaussian multi-mode states play an important role in many
applications, most notably in continuous variable quantum information.
We showed that the normal parameters of an $S$-mode Gaussian, which
consist of the eigenvalues of the state's covariance matrix and
associated absolute displacements, can be inferred from the first
$8S+1$ total photon-number probabilities, which in turn can be
inferred from typical photon-number resolving measurements, even
inefficient or approximate ones. This implies that a simple
configuration involving only a mode-insensitive detector can be used
to monitor Gaussian states. We obtained an efficient algebraic method
to infer the normal parameters from the photon-number
probabilities. Further work is required to determine sensitivity to
statistical estimation error and the number of samples required to
obtain sufficiently precise information about the parameters in any
given context. The normal parameters consist of at most \(4S\) real
quantities, so at least \(4S\) photon number probabilities are
required to infer them. An interesting problem is to determine whether
\(4S\) photon number probabilities suffice. More generally, for which
sets \(J\) of natural numbers can the normal parameters be determined
from the photon number probabilities \(p_{n}\) for \(n\in J\)?
Measurement configurations involving multiple mode-sensitive detectors
and local oscillators can give substantially more information at the
cost of additional complexity. Our work shows that important
parameters of Gaussian states can be obtained from a small number of
estimated quantities with a simple experimental configuration without
requiring fast, precise or especially efficient detectors.
 
\begin{acknowledgments}
  This work includes contributions of the National Institute of
  Standards and Technology, which are not subject to U.S. copyright.
  A. Avagyan acknowledges support from the Professional Research Experience Program (PREP) operated jointly by NIST and the University of Colorado.
\end{acknowledgments}

\bibliography{gg_prl.bib}

\begin{thebibliography}{33}%
\makeatletter
\providecommand \@ifxundefined [1]{%
 \@ifx{#1\undefined}
}%
\providecommand \@ifnum [1]{%
 \ifnum #1\expandafter \@firstoftwo
 \else \expandafter \@secondoftwo
 \fi
}%
\providecommand \@ifx [1]{%
 \ifx #1\expandafter \@firstoftwo
 \else \expandafter \@secondoftwo
 \fi
}%
\providecommand \natexlab [1]{#1}%
\providecommand \enquote  [1]{``#1''}%
\providecommand \bibnamefont  [1]{#1}%
\providecommand \bibfnamefont [1]{#1}%
\providecommand \citenamefont [1]{#1}%
\providecommand \href@noop [0]{\@secondoftwo}%
\providecommand \href [0]{\begingroup \@sanitize@url \@href}%
\providecommand \@href[1]{\@@startlink{#1}\@@href}%
\providecommand \@@href[1]{\endgroup#1\@@endlink}%
\providecommand \@sanitize@url [0]{\catcode `\\12\catcode `\$12\catcode
  `\&12\catcode `\#12\catcode `\^12\catcode `\_12\catcode `\%12\relax}%
\providecommand \@@startlink[1]{}%
\providecommand \@@endlink[0]{}%
\providecommand \url  [0]{\begingroup\@sanitize@url \@url }%
\providecommand \@url [1]{\endgroup\@href {#1}{\urlprefix }}%
\providecommand \urlprefix  [0]{URL }%
\providecommand \Eprint [0]{\href }%
\providecommand \doibase [0]{https://doi.org/}%
\providecommand \selectlanguage [0]{\@gobble}%
\providecommand \bibinfo  [0]{\@secondoftwo}%
\providecommand \bibfield  [0]{\@secondoftwo}%
\providecommand \translation [1]{[#1]}%
\providecommand \BibitemOpen [0]{}%
\providecommand \bibitemStop [0]{}%
\providecommand \bibitemNoStop [0]{.\EOS\space}%
\providecommand \EOS [0]{\spacefactor3000\relax}%
\providecommand \BibitemShut  [1]{\csname bibitem#1\endcsname}%
\let\auto@bib@innerbib\@empty
\bibitem [{\citenamefont {Wang}\ \emph {et~al.}(2007)\citenamefont {Wang},
  \citenamefont {Hiroshima}, \citenamefont {Tomita},\ and\ \citenamefont
  {Hayashi}}]{wang2007quantum}%
  \BibitemOpen
  \bibfield  {author} {\bibinfo {author} {\bibfnamefont {X.-B.}\ \bibnamefont
  {Wang}}, \bibinfo {author} {\bibfnamefont {T.}~\bibnamefont {Hiroshima}},
  \bibinfo {author} {\bibfnamefont {A.}~\bibnamefont {Tomita}},\ and\ \bibinfo
  {author} {\bibfnamefont {M.}~\bibnamefont {Hayashi}},\ }\bibfield  {title}
  {\bibinfo {title} {Quantum information with {G}aussian states},\ }\href
  {https://doi.org/https://doi.org/10.1016/j.physrep.2007.04.005} {\bibfield
  {journal} {\bibinfo  {journal} {Physics Reports}\ }\textbf {\bibinfo {volume}
  {448}},\ \bibinfo {pages} {1} (\bibinfo {year} {2007})}\BibitemShut {NoStop}%
\bibitem [{\citenamefont {Weedbrook}\ \emph {et~al.}(2012)\citenamefont
  {Weedbrook}, \citenamefont {Pirandola}, \citenamefont {Garcia-Patron},
  \citenamefont {Cerf}, \citenamefont {Ralph}, \citenamefont {Shapiro},\ and\
  \citenamefont {Lloyd}}]{weedbrook:qc2012a}%
  \BibitemOpen
  \bibfield  {author} {\bibinfo {author} {\bibfnamefont {C.}~\bibnamefont
  {Weedbrook}}, \bibinfo {author} {\bibfnamefont {S.}~\bibnamefont
  {Pirandola}}, \bibinfo {author} {\bibfnamefont {R.}~\bibnamefont
  {Garcia-Patron}}, \bibinfo {author} {\bibfnamefont {N.~J.}\ \bibnamefont
  {Cerf}}, \bibinfo {author} {\bibfnamefont {T.~C.}\ \bibnamefont {Ralph}},
  \bibinfo {author} {\bibfnamefont {J.~H.}\ \bibnamefont {Shapiro}},\ and\
  \bibinfo {author} {\bibfnamefont {S.}~\bibnamefont {Lloyd}},\ }\bibfield
  {title} {\bibinfo {title} {Gaussian quantum information},\ }\href
  {https://doi.org/10.1103/RevModPhys.84.621} {\bibfield  {journal} {\bibinfo
  {journal} {Rev. Mod. Phys.}\ }\textbf {\bibinfo {volume} {84}},\ \bibinfo
  {pages} {621} (\bibinfo {year} {2012})}\BibitemShut {NoStop}%
\bibitem [{\citenamefont {Paris}\ \emph {et~al.}(2003)\citenamefont {Paris},
  \citenamefont {Illuminati}, \citenamefont {Serafini},\ and\ \citenamefont
  {De~Siena}}]{paris2003purity}%
  \BibitemOpen
  \bibfield  {author} {\bibinfo {author} {\bibfnamefont {M.~G.~A.}\
  \bibnamefont {Paris}}, \bibinfo {author} {\bibfnamefont {F.}~\bibnamefont
  {Illuminati}}, \bibinfo {author} {\bibfnamefont {A.}~\bibnamefont
  {Serafini}},\ and\ \bibinfo {author} {\bibfnamefont {S.}~\bibnamefont
  {De~Siena}},\ }\bibfield  {title} {\bibinfo {title} {Purity of {G}aussian
  states: Measurement schemes and time evolution in noisy channels},\ }\href
  {https://doi.org/10.1103/PhysRevA.68.012314} {\bibfield  {journal} {\bibinfo
  {journal} {Phys. Rev. A}\ }\textbf {\bibinfo {volume} {68}},\ \bibinfo
  {pages} {012314} (\bibinfo {year} {2003})}\BibitemShut {NoStop}%
\bibitem [{\citenamefont {Laurat}\ \emph {et~al.}(2005)\citenamefont {Laurat},
  \citenamefont {Keller}, \citenamefont {Oliveira-Huguenin}, \citenamefont
  {Fabre}, \citenamefont {Coudreau}, \citenamefont {Serafini}, \citenamefont
  {Adesso},\ and\ \citenamefont {Illuminati}}]{laurat2005entanglement}%
  \BibitemOpen
  \bibfield  {author} {\bibinfo {author} {\bibfnamefont {J.}~\bibnamefont
  {Laurat}}, \bibinfo {author} {\bibfnamefont {G.}~\bibnamefont {Keller}},
  \bibinfo {author} {\bibfnamefont {J.~A.}\ \bibnamefont {Oliveira-Huguenin}},
  \bibinfo {author} {\bibfnamefont {C.}~\bibnamefont {Fabre}}, \bibinfo
  {author} {\bibfnamefont {T.}~\bibnamefont {Coudreau}}, \bibinfo {author}
  {\bibfnamefont {A.}~\bibnamefont {Serafini}}, \bibinfo {author}
  {\bibfnamefont {G.}~\bibnamefont {Adesso}},\ and\ \bibinfo {author}
  {\bibfnamefont {F.}~\bibnamefont {Illuminati}},\ }\bibfield  {title}
  {\bibinfo {title} {Entanglement of two-mode {G}aussian states:
  Characterization and experimental production and manipulation},\ }\href
  {https://doi.org/10.1088/1464-4266/7/12/021} {\bibfield  {journal} {\bibinfo
  {journal} {Journal of Optics B: Quantum and Semiclassical Optics}\ }\textbf
  {\bibinfo {volume} {7}},\ \bibinfo {pages} {S577} (\bibinfo {year}
  {2005})}\BibitemShut {NoStop}%
\bibitem [{\citenamefont {D'Auria}\ \emph {et~al.}(2005)\citenamefont
  {D'Auria}, \citenamefont {Porzio}, \citenamefont {Solimeno}, \citenamefont
  {Olivares},\ and\ \citenamefont {Paris}}]{dauria2005characterization}%
  \BibitemOpen
  \bibfield  {author} {\bibinfo {author} {\bibfnamefont {V.}~\bibnamefont
  {D'Auria}}, \bibinfo {author} {\bibfnamefont {A.}~\bibnamefont {Porzio}},
  \bibinfo {author} {\bibfnamefont {S.}~\bibnamefont {Solimeno}}, \bibinfo
  {author} {\bibfnamefont {S.}~\bibnamefont {Olivares}},\ and\ \bibinfo
  {author} {\bibfnamefont {M.~G.~A.}\ \bibnamefont {Paris}},\ }\bibfield
  {title} {\bibinfo {title} {Characterization of bipartite states using a
  single homodyne detector},\ }\href
  {https://doi.org/10.1088/1464-4266/7/12/044} {\bibfield  {journal} {\bibinfo
  {journal} {Journal of Optics B: Quantum and Semiclassical Optics}\ }\textbf
  {\bibinfo {volume} {7}},\ \bibinfo {pages} {S750} (\bibinfo {year}
  {2005})}\BibitemShut {NoStop}%
\bibitem [{\citenamefont {Porzio}\ \emph {et~al.}(2007)\citenamefont {Porzio},
  \citenamefont {D'auria}, \citenamefont {Solimeno}, \citenamefont {Olivares},\
  and\ \citenamefont {Paris}}]{porzio2007homodyne}%
  \BibitemOpen
  \bibfield  {author} {\bibinfo {author} {\bibfnamefont {A.}~\bibnamefont
  {Porzio}}, \bibinfo {author} {\bibfnamefont {V.}~\bibnamefont {D'auria}},
  \bibinfo {author} {\bibfnamefont {S.}~\bibnamefont {Solimeno}}, \bibinfo
  {author} {\bibfnamefont {S.}~\bibnamefont {Olivares}},\ and\ \bibinfo
  {author} {\bibfnamefont {M.~G.~A.}\ \bibnamefont {Paris}},\ }\bibfield
  {title} {\bibinfo {title} {Homodyne characterization of continuous variable
  bipartite states},\ }\href {https://doi.org/10.1142/S0219749907002529}
  {\bibfield  {journal} {\bibinfo  {journal} {International Journal of Quantum
  Information}\ }\textbf {\bibinfo {volume} {05}},\ \bibinfo {pages} {63}
  (\bibinfo {year} {2007})}\BibitemShut {NoStop}%
\bibitem [{\citenamefont {\ifmmode \check{R}\else
  \v{R}\fi{}eh\'a\ifmmode~\check{c}\else \v{c}\fi{}ek}\ \emph
  {et~al.}(2009)\citenamefont {\ifmmode \check{R}\else
  \v{R}\fi{}eh\'a\ifmmode~\check{c}\else \v{c}\fi{}ek}, \citenamefont
  {Olivares}, \citenamefont {Mogilevtsev}, \citenamefont {Hradil},
  \citenamefont {Paris}, \citenamefont {Fornaro}, \citenamefont {D'Auria},
  \citenamefont {Porzio},\ and\ \citenamefont
  {Solimeno}}]{rehacek2009effective}%
  \BibitemOpen
  \bibfield  {author} {\bibinfo {author} {\bibfnamefont {J.}~\bibnamefont
  {\ifmmode \check{R}\else \v{R}\fi{}eh\'a\ifmmode~\check{c}\else
  \v{c}\fi{}ek}}, \bibinfo {author} {\bibfnamefont {S.}~\bibnamefont
  {Olivares}}, \bibinfo {author} {\bibfnamefont {D.}~\bibnamefont
  {Mogilevtsev}}, \bibinfo {author} {\bibfnamefont {Z.}~\bibnamefont {Hradil}},
  \bibinfo {author} {\bibfnamefont {M.~G.~A.}\ \bibnamefont {Paris}}, \bibinfo
  {author} {\bibfnamefont {S.}~\bibnamefont {Fornaro}}, \bibinfo {author}
  {\bibfnamefont {V.}~\bibnamefont {D'Auria}}, \bibinfo {author} {\bibfnamefont
  {A.}~\bibnamefont {Porzio}},\ and\ \bibinfo {author} {\bibfnamefont
  {S.}~\bibnamefont {Solimeno}},\ }\bibfield  {title} {\bibinfo {title}
  {Effective method to estimate multidimensional {G}aussian states},\ }\href
  {https://doi.org/10.1103/PhysRevA.79.032111} {\bibfield  {journal} {\bibinfo
  {journal} {Phys. Rev. A}\ }\textbf {\bibinfo {volume} {79}},\ \bibinfo
  {pages} {032111} (\bibinfo {year} {2009})}\BibitemShut {NoStop}%
\bibitem [{\citenamefont {D'Auria}\ \emph {et~al.}(2009)\citenamefont
  {D'Auria}, \citenamefont {Fornaro}, \citenamefont {Porzio}, \citenamefont
  {Solimeno}, \citenamefont {Olivares},\ and\ \citenamefont
  {Paris}}]{dauria2009full}%
  \BibitemOpen
  \bibfield  {author} {\bibinfo {author} {\bibfnamefont {V.}~\bibnamefont
  {D'Auria}}, \bibinfo {author} {\bibfnamefont {S.}~\bibnamefont {Fornaro}},
  \bibinfo {author} {\bibfnamefont {A.}~\bibnamefont {Porzio}}, \bibinfo
  {author} {\bibfnamefont {S.}~\bibnamefont {Solimeno}}, \bibinfo {author}
  {\bibfnamefont {S.}~\bibnamefont {Olivares}},\ and\ \bibinfo {author}
  {\bibfnamefont {M.~G.~A.}\ \bibnamefont {Paris}},\ }\bibfield  {title}
  {\bibinfo {title} {Full characterization of {G}aussian bipartite entangled
  states by a single homodyne detector},\ }\href
  {https://doi.org/10.1103/PhysRevLett.102.020502} {\bibfield  {journal}
  {\bibinfo  {journal} {Phys. Rev. Lett.}\ }\textbf {\bibinfo {volume} {102}},\
  \bibinfo {pages} {020502} (\bibinfo {year} {2009})}\BibitemShut {NoStop}%
\bibitem [{\citenamefont {Paternostro}\ \emph {et~al.}(2009)\citenamefont
  {Paternostro}, \citenamefont {Jeong},\ and\ \citenamefont
  {Ralph}}]{paternostro:qc2009a}%
  \BibitemOpen
  \bibfield  {author} {\bibinfo {author} {\bibfnamefont {M.}~\bibnamefont
  {Paternostro}}, \bibinfo {author} {\bibfnamefont {H.}~\bibnamefont {Jeong}},\
  and\ \bibinfo {author} {\bibfnamefont {T.~C.}\ \bibnamefont {Ralph}},\
  }\bibfield  {title} {\bibinfo {title} {Violations of {Bell}’s inequality
  for {Gaussian} states with homodyne detection and nonlinear interactions},\
  }\href {https://doi.org/10.1103/PhysRevA.79.012101} {\bibfield  {journal}
  {\bibinfo  {journal} {Phys. Rev. A}\ }\textbf {\bibinfo {volume} {79}},\
  \bibinfo {pages} {012101/1} (\bibinfo {year} {2009})}\BibitemShut {NoStop}%
\bibitem [{\citenamefont {Buono}\ \emph {et~al.}(2010)\citenamefont {Buono},
  \citenamefont {Nocerino}, \citenamefont {D'Auria}, \citenamefont {Porzio},
  \citenamefont {Olivares},\ and\ \citenamefont {Paris}}]{buono2010quantum}%
  \BibitemOpen
  \bibfield  {author} {\bibinfo {author} {\bibfnamefont {D.}~\bibnamefont
  {Buono}}, \bibinfo {author} {\bibfnamefont {G.}~\bibnamefont {Nocerino}},
  \bibinfo {author} {\bibfnamefont {V.}~\bibnamefont {D'Auria}}, \bibinfo
  {author} {\bibfnamefont {A.}~\bibnamefont {Porzio}}, \bibinfo {author}
  {\bibfnamefont {S.}~\bibnamefont {Olivares}},\ and\ \bibinfo {author}
  {\bibfnamefont {M.~G.~A.}\ \bibnamefont {Paris}},\ }\bibfield  {title}
  {\bibinfo {title} {Quantum characterization of bipartite {G}aussian states},\
  }\href {http://josab.osa.org/abstract.cfm?URI=josab-27-6-A110} {\bibfield
  {journal} {\bibinfo  {journal} {J. Opt. Soc. Am. B}\ }\textbf {\bibinfo
  {volume} {27}},\ \bibinfo {pages} {A110} (\bibinfo {year}
  {2010})}\BibitemShut {NoStop}%
\bibitem [{\citenamefont {Blandino}\ \emph {et~al.}(2012)\citenamefont
  {Blandino}, \citenamefont {Genoni}, \citenamefont {Etesse}, \citenamefont
  {Barbieri}, \citenamefont {Paris}, \citenamefont {Grangier},\ and\
  \citenamefont {Tualle-Brouri}}]{blandino:qc2012a}%
  \BibitemOpen
  \bibfield  {author} {\bibinfo {author} {\bibfnamefont {R.}~\bibnamefont
  {Blandino}}, \bibinfo {author} {\bibfnamefont {M.~G.}\ \bibnamefont
  {Genoni}}, \bibinfo {author} {\bibfnamefont {J.}~\bibnamefont {Etesse}},
  \bibinfo {author} {\bibfnamefont {M.}~\bibnamefont {Barbieri}}, \bibinfo
  {author} {\bibfnamefont {M.~G.~A.}\ \bibnamefont {Paris}}, \bibinfo {author}
  {\bibfnamefont {P.}~\bibnamefont {Grangier}},\ and\ \bibinfo {author}
  {\bibfnamefont {R.}~\bibnamefont {Tualle-Brouri}},\ }\bibfield  {title}
  {\bibinfo {title} {Homodyne estimation of {G}aussian quantum discord},\
  }\href {https://doi.org/10.1103/PhysRevLett.109.180402} {\bibfield  {journal}
  {\bibinfo  {journal} {Phys. Rev. Lett.}\ }\textbf {\bibinfo {volume} {109}},\
  \bibinfo {pages} {180402} (\bibinfo {year} {2012})}\BibitemShut {NoStop}%
\bibitem [{\citenamefont {Esposito}\ \emph {et~al.}(2014)\citenamefont
  {Esposito}, \citenamefont {Benatti}, \citenamefont {Floreanini},
  \citenamefont {Olivares}, \citenamefont {Randi}, \citenamefont {Titimbo},
  \citenamefont {Pividori}, \citenamefont {Novelli}, \citenamefont {Cilento},
  \citenamefont {Parmigiani},\ and\ \citenamefont
  {Fausti}}]{esposito2014pulsed}%
  \BibitemOpen
  \bibfield  {author} {\bibinfo {author} {\bibfnamefont {M.}~\bibnamefont
  {Esposito}}, \bibinfo {author} {\bibfnamefont {F.}~\bibnamefont {Benatti}},
  \bibinfo {author} {\bibfnamefont {R.}~\bibnamefont {Floreanini}}, \bibinfo
  {author} {\bibfnamefont {S.}~\bibnamefont {Olivares}}, \bibinfo {author}
  {\bibfnamefont {F.}~\bibnamefont {Randi}}, \bibinfo {author} {\bibfnamefont
  {K.}~\bibnamefont {Titimbo}}, \bibinfo {author} {\bibfnamefont
  {M.}~\bibnamefont {Pividori}}, \bibinfo {author} {\bibfnamefont
  {F.}~\bibnamefont {Novelli}}, \bibinfo {author} {\bibfnamefont
  {F.}~\bibnamefont {Cilento}}, \bibinfo {author} {\bibfnamefont
  {F.}~\bibnamefont {Parmigiani}},\ and\ \bibinfo {author} {\bibfnamefont
  {D.}~\bibnamefont {Fausti}},\ }\bibfield  {title} {\bibinfo {title} {Pulsed
  homodyne {G}aussian quantum tomography with low detection efficiency},\
  }\href {https://doi.org/10.1088/1367-2630/16/4/043004} {\bibfield  {journal}
  {\bibinfo  {journal} {New Journal of Physics}\ }\textbf {\bibinfo {volume}
  {16}},\ \bibinfo {pages} {043004} (\bibinfo {year} {2014})}\BibitemShut
  {NoStop}%
\bibitem [{\citenamefont {Fiur\'a\ifmmode~\check{s}\else \v{s}\fi{}ek}\ and\
  \citenamefont {Cerf}(2004)}]{fiurasek2004how}%
  \BibitemOpen
  \bibfield  {author} {\bibinfo {author} {\bibfnamefont {J.}~\bibnamefont
  {Fiur\'a\ifmmode~\check{s}\else \v{s}\fi{}ek}}\ and\ \bibinfo {author}
  {\bibfnamefont {N.~J.}\ \bibnamefont {Cerf}},\ }\bibfield  {title} {\bibinfo
  {title} {How to measure squeezing and entanglement of {G}aussian states
  without homodyning},\ }\href {https://doi.org/10.1103/PhysRevLett.93.063601}
  {\bibfield  {journal} {\bibinfo  {journal} {Phys. Rev. Lett.}\ }\textbf
  {\bibinfo {volume} {93}},\ \bibinfo {pages} {063601} (\bibinfo {year}
  {2004})}\BibitemShut {NoStop}%
\bibitem [{\citenamefont {Wenger}\ \emph {et~al.}(2004)\citenamefont {Wenger},
  \citenamefont {Fiur\'a\ifmmode~\check{s}\else \v{s}\fi{}ek}, \citenamefont
  {Tualle-Brouri}, \citenamefont {Cerf},\ and\ \citenamefont
  {Grangier}}]{wenger2004pulsed}%
  \BibitemOpen
  \bibfield  {author} {\bibinfo {author} {\bibfnamefont {J.}~\bibnamefont
  {Wenger}}, \bibinfo {author} {\bibfnamefont {J.}~\bibnamefont
  {Fiur\'a\ifmmode~\check{s}\else \v{s}\fi{}ek}}, \bibinfo {author}
  {\bibfnamefont {R.}~\bibnamefont {Tualle-Brouri}}, \bibinfo {author}
  {\bibfnamefont {N.~J.}\ \bibnamefont {Cerf}},\ and\ \bibinfo {author}
  {\bibfnamefont {P.}~\bibnamefont {Grangier}},\ }\bibfield  {title} {\bibinfo
  {title} {Pulsed squeezed vacuum measurements without homodyning},\ }\href
  {https://doi.org/10.1103/PhysRevA.70.053812} {\bibfield  {journal} {\bibinfo
  {journal} {Phys. Rev. A}\ }\textbf {\bibinfo {volume} {70}},\ \bibinfo
  {pages} {053812} (\bibinfo {year} {2004})}\BibitemShut {NoStop}%
\bibitem [{\citenamefont {Wallentowitz}\ and\ \citenamefont
  {Vogel}(1996)}]{wallentowitz1996unbalanced}%
  \BibitemOpen
  \bibfield  {author} {\bibinfo {author} {\bibfnamefont {S.}~\bibnamefont
  {Wallentowitz}}\ and\ \bibinfo {author} {\bibfnamefont {W.}~\bibnamefont
  {Vogel}},\ }\bibfield  {title} {\bibinfo {title} {Unbalanced homodyning for
  quantum state measurements},\ }\href
  {https://doi.org/10.1103/PhysRevA.53.4528} {\bibfield  {journal} {\bibinfo
  {journal} {Phys. Rev. A}\ }\textbf {\bibinfo {volume} {53}},\ \bibinfo
  {pages} {4528} (\bibinfo {year} {1996})}\BibitemShut {NoStop}%
\bibitem [{\citenamefont {Banaszek}\ and\ \citenamefont
  {W\'odkiewicz}(1996)}]{konrad1996direct}%
  \BibitemOpen
  \bibfield  {author} {\bibinfo {author} {\bibfnamefont {K.}~\bibnamefont
  {Banaszek}}\ and\ \bibinfo {author} {\bibfnamefont {K.}~\bibnamefont
  {W\'odkiewicz}},\ }\bibfield  {title} {\bibinfo {title} {Direct probing of
  quantum phase space by photon counting},\ }\href
  {https://doi.org/10.1103/PhysRevLett.76.4344} {\bibfield  {journal} {\bibinfo
   {journal} {Phys. Rev. Lett.}\ }\textbf {\bibinfo {volume} {76}},\ \bibinfo
  {pages} {4344} (\bibinfo {year} {1996})}\BibitemShut {NoStop}%
\bibitem [{\citenamefont {Man'ko}\ and\ \citenamefont
  {Man'ko}(2003)}]{manko2003photon}%
  \BibitemOpen
  \bibfield  {author} {\bibinfo {author} {\bibfnamefont {O.}~\bibnamefont
  {Man'ko}}\ and\ \bibinfo {author} {\bibfnamefont {V.~I.}\ \bibnamefont
  {Man'ko}},\ }\bibfield  {title} {\bibinfo {title} {Photon-number tomography
  of multimode states and positivity of the density matrix},\ }\href
  {https://doi.org/10.1023/A:1025876210639} {\bibfield  {journal} {\bibinfo
  {journal} {Journal of Russian Laser Research}\ }\textbf {\bibinfo {volume}
  {24}},\ \bibinfo {pages} {497} (\bibinfo {year} {2003})}\BibitemShut
  {NoStop}%
\bibitem [{\citenamefont {Man'ko}\ and\ \citenamefont
  {Man'ko}(2009)}]{manko2009photon}%
  \BibitemOpen
  \bibfield  {author} {\bibinfo {author} {\bibfnamefont {O.~V.}\ \bibnamefont
  {Man'ko}}\ and\ \bibinfo {author} {\bibfnamefont {V.~I.}\ \bibnamefont
  {Man'ko}},\ }\bibfield  {title} {\bibinfo {title} {Photon number and optical
  tomograms for {G}aussian states},\ }\href
  {https://doi.org/10.1134/S1054660X09150286} {\bibfield  {journal} {\bibinfo
  {journal} {Laser Physics}\ }\textbf {\bibinfo {volume} {19}},\ \bibinfo
  {pages} {1804} (\bibinfo {year} {2009})}\BibitemShut {NoStop}%
\bibitem [{\citenamefont {Leibfried}\ \emph {et~al.}(1996)\citenamefont
  {Leibfried}, \citenamefont {Meekhof}, \citenamefont {King}, \citenamefont
  {Monroe}, \citenamefont {Itano},\ and\ \citenamefont
  {Wineland}}]{leibfried1996exp}%
  \BibitemOpen
  \bibfield  {author} {\bibinfo {author} {\bibfnamefont {D.}~\bibnamefont
  {Leibfried}}, \bibinfo {author} {\bibfnamefont {D.~M.}\ \bibnamefont
  {Meekhof}}, \bibinfo {author} {\bibfnamefont {B.~E.}\ \bibnamefont {King}},
  \bibinfo {author} {\bibfnamefont {C.}~\bibnamefont {Monroe}}, \bibinfo
  {author} {\bibfnamefont {W.~M.}\ \bibnamefont {Itano}},\ and\ \bibinfo
  {author} {\bibfnamefont {D.~J.}\ \bibnamefont {Wineland}},\ }\bibfield
  {title} {\bibinfo {title} {Experimental determination of the motional quantum
  state of a trapped atom},\ }\href
  {https://doi.org/10.1103/PhysRevLett.77.4281} {\bibfield  {journal} {\bibinfo
   {journal} {Phys. Rev. Lett.}\ }\textbf {\bibinfo {volume} {77}},\ \bibinfo
  {pages} {4281} (\bibinfo {year} {1996})}\BibitemShut {NoStop}%
\bibitem [{\citenamefont {Nogues}\ \emph {et~al.}(2000)\citenamefont {Nogues},
  \citenamefont {Rauschenbeutel}, \citenamefont {Osnaghi}, \citenamefont
  {Bertet}, \citenamefont {Brune}, \citenamefont {Raimond}, \citenamefont
  {Haroche}, \citenamefont {Lutterbach},\ and\ \citenamefont
  {Davidovich}}]{nogues2000meas}%
  \BibitemOpen
  \bibfield  {author} {\bibinfo {author} {\bibfnamefont {G.}~\bibnamefont
  {Nogues}}, \bibinfo {author} {\bibfnamefont {A.}~\bibnamefont
  {Rauschenbeutel}}, \bibinfo {author} {\bibfnamefont {S.}~\bibnamefont
  {Osnaghi}}, \bibinfo {author} {\bibfnamefont {P.}~\bibnamefont {Bertet}},
  \bibinfo {author} {\bibfnamefont {M.}~\bibnamefont {Brune}}, \bibinfo
  {author} {\bibfnamefont {J.~M.}\ \bibnamefont {Raimond}}, \bibinfo {author}
  {\bibfnamefont {S.}~\bibnamefont {Haroche}}, \bibinfo {author} {\bibfnamefont
  {L.~G.}\ \bibnamefont {Lutterbach}},\ and\ \bibinfo {author} {\bibfnamefont
  {L.}~\bibnamefont {Davidovich}},\ }\bibfield  {title} {\bibinfo {title}
  {Measurement of a negative value for the {W}igner function of radiation},\
  }\href {https://doi.org/10.1103/PhysRevA.62.054101} {\bibfield  {journal}
  {\bibinfo  {journal} {Phys. Rev. A}\ }\textbf {\bibinfo {volume} {62}},\
  \bibinfo {pages} {054101} (\bibinfo {year} {2000})}\BibitemShut {NoStop}%
\bibitem [{\citenamefont {Wu}\ \emph {et~al.}(2019)\citenamefont {Wu},
  \citenamefont {Toda},\ and\ \citenamefont {Hofmann}}]{wu2019quantum}%
  \BibitemOpen
  \bibfield  {author} {\bibinfo {author} {\bibfnamefont {J.-Y.}\ \bibnamefont
  {Wu}}, \bibinfo {author} {\bibfnamefont {N.}~\bibnamefont {Toda}},\ and\
  \bibinfo {author} {\bibfnamefont {H.~F.}\ \bibnamefont {Hofmann}},\
  }\bibfield  {title} {\bibinfo {title} {Quantum enhancement of sensitivity
  achieved by photon-number-resolving detection in the dark port of a two-path
  interferometer operating at high intensities},\ }\href
  {https://doi.org/10.1103/PhysRevA.100.013814} {\bibfield  {journal} {\bibinfo
   {journal} {Phys. Rev. A}\ }\textbf {\bibinfo {volume} {100}},\ \bibinfo
  {pages} {013814} (\bibinfo {year} {2019})}\BibitemShut {NoStop}%
\bibitem [{\citenamefont {Burd}\ \emph {et~al.}(2019)\citenamefont {Burd},
  \citenamefont {Srinivas}, \citenamefont {Bollinger}, \citenamefont {Wilson},
  \citenamefont {Wineland}, \citenamefont {Leibfried}, \citenamefont
  {Slichter},\ and\ \citenamefont {Allcock}}]{burd2019quantum}%
  \BibitemOpen
  \bibfield  {author} {\bibinfo {author} {\bibfnamefont {S.~C.}\ \bibnamefont
  {Burd}}, \bibinfo {author} {\bibfnamefont {R.}~\bibnamefont {Srinivas}},
  \bibinfo {author} {\bibfnamefont {J.~J.}\ \bibnamefont {Bollinger}}, \bibinfo
  {author} {\bibfnamefont {A.~C.}\ \bibnamefont {Wilson}}, \bibinfo {author}
  {\bibfnamefont {D.~J.}\ \bibnamefont {Wineland}}, \bibinfo {author}
  {\bibfnamefont {D.}~\bibnamefont {Leibfried}}, \bibinfo {author}
  {\bibfnamefont {D.~H.}\ \bibnamefont {Slichter}},\ and\ \bibinfo {author}
  {\bibfnamefont {D.~T.~C.}\ \bibnamefont {Allcock}},\ }\bibfield  {title}
  {\bibinfo {title} {Quantum amplification of mechanical oscillator motion},\
  }\href {https://doi.org/10.1126/science.aaw2884} {\bibfield  {journal}
  {\bibinfo  {journal} {Science}\ }\textbf {\bibinfo {volume} {364}},\ \bibinfo
  {pages} {1163} (\bibinfo {year} {2019})}\BibitemShut {NoStop}%
\bibitem [{\citenamefont {Burenkov}\ \emph {et~al.}(2017)\citenamefont
  {Burenkov}, \citenamefont {Sharma}, \citenamefont {Gerrits}, \citenamefont
  {Harder}, \citenamefont {Bartley}, \citenamefont {Silberhorn}, \citenamefont
  {Goldschmidt},\ and\ \citenamefont {Polyakov}}]{burenkov2017full}%
  \BibitemOpen
  \bibfield  {author} {\bibinfo {author} {\bibfnamefont {I.~A.}\ \bibnamefont
  {Burenkov}}, \bibinfo {author} {\bibfnamefont {A.~K.}\ \bibnamefont
  {Sharma}}, \bibinfo {author} {\bibfnamefont {T.}~\bibnamefont {Gerrits}},
  \bibinfo {author} {\bibfnamefont {G.}~\bibnamefont {Harder}}, \bibinfo
  {author} {\bibfnamefont {T.~J.}\ \bibnamefont {Bartley}}, \bibinfo {author}
  {\bibfnamefont {C.}~\bibnamefont {Silberhorn}}, \bibinfo {author}
  {\bibfnamefont {E.~A.}\ \bibnamefont {Goldschmidt}},\ and\ \bibinfo {author}
  {\bibfnamefont {S.~V.}\ \bibnamefont {Polyakov}},\ }\bibfield  {title}
  {\bibinfo {title} {Full statistical mode reconstruction of a light field via
  a photon-number-resolved measurement},\ }\href
  {https://doi.org/10.1103/PhysRevA.95.053806} {\bibfield  {journal} {\bibinfo
  {journal} {Phys. Rev. A}\ }\textbf {\bibinfo {volume} {95}},\ \bibinfo
  {pages} {053806} (\bibinfo {year} {2017})}\BibitemShut {NoStop}%
\bibitem [{\citenamefont {Zuo}\ \emph {et~al.}(2022)\citenamefont {Zuo},
  \citenamefont {Zhang}, \citenamefont {Li}, \citenamefont {Zhu}, \citenamefont
  {Guo},\ and\ \citenamefont {Zhang}}]{zuo2022determination}%
  \BibitemOpen
  \bibfield  {author} {\bibinfo {author} {\bibfnamefont {G.}~\bibnamefont
  {Zuo}}, \bibinfo {author} {\bibfnamefont {Y.}~\bibnamefont {Zhang}}, \bibinfo
  {author} {\bibfnamefont {J.}~\bibnamefont {Li}}, \bibinfo {author}
  {\bibfnamefont {S.}~\bibnamefont {Zhu}}, \bibinfo {author} {\bibfnamefont
  {Y.}~\bibnamefont {Guo}},\ and\ \bibinfo {author} {\bibfnamefont
  {T.}~\bibnamefont {Zhang}},\ }\bibfield  {title} {\bibinfo {title}
  {Determination of weakly squeezed vacuum states through photon statistics
  measurement},\ }\href
  {https://doi.org/https://doi.org/10.1016/j.physleta.2022.128133} {\bibfield
  {journal} {\bibinfo  {journal} {Physics Letters A}\ }\textbf {\bibinfo
  {volume} {439}},\ \bibinfo {pages} {128133} (\bibinfo {year}
  {2022})}\BibitemShut {NoStop}%
\bibitem [{\citenamefont {Bezerra}\ \emph {et~al.}(2021)\citenamefont
  {Bezerra}, \citenamefont {Vasconcelos},\ and\ \citenamefont
  {Glancy}}]{bezerra2021quadrature}%
  \BibitemOpen
  \bibfield  {author} {\bibinfo {author} {\bibfnamefont {I.~P.}\ \bibnamefont
  {Bezerra}}, \bibinfo {author} {\bibfnamefont {H.~M.}\ \bibnamefont
  {Vasconcelos}},\ and\ \bibinfo {author} {\bibfnamefont {S.}~\bibnamefont
  {Glancy}},\ }\href {https://doi.org/10.48550/ARXIV.2111.04923} {\bibinfo
  {title} {Quadrature squeezing and temperature estimation from the {F}ock
  distribution}} (\bibinfo {year} {2021})\BibitemShut {NoStop}%
\bibitem [{\citenamefont {Parthasarathy}\ and\ \citenamefont
  {Sengupta}(2015)}]{parthasarathy2015from}%
  \BibitemOpen
  \bibfield  {author} {\bibinfo {author} {\bibfnamefont {K.~R.}\ \bibnamefont
  {Parthasarathy}}\ and\ \bibinfo {author} {\bibfnamefont {R.}~\bibnamefont
  {Sengupta}},\ }\bibfield  {title} {\bibinfo {title} {From particle counting
  to {G}aussian tomography},\ }\href
  {https://doi.org/10.1142/S021902571550023X} {\bibfield  {journal} {\bibinfo
  {journal} {Infinite Dimensional Analysis, Quantum Probability and Related
  Topics}\ }\textbf {\bibinfo {volume} {18}},\ \bibinfo {pages} {1550023}
  (\bibinfo {year} {2015})}\BibitemShut {NoStop}%
\bibitem [{\citenamefont {Kumar}\ \emph {et~al.}(2020)\citenamefont {Kumar},
  \citenamefont {Sengupta},\ and\ \citenamefont {Arvind}}]{kumar2020optimal}%
  \BibitemOpen
  \bibfield  {author} {\bibinfo {author} {\bibfnamefont {C.}~\bibnamefont
  {Kumar}}, \bibinfo {author} {\bibfnamefont {R.}~\bibnamefont {Sengupta}},\
  and\ \bibinfo {author} {\bibnamefont {Arvind}},\ }\bibfield  {title}
  {\bibinfo {title} {Optimal characterization of {G}aussian channels using
  photon-number-resolving detectors},\ }\href
  {https://doi.org/10.1103/PhysRevA.102.012616} {\bibfield  {journal} {\bibinfo
   {journal} {Phys. Rev. A}\ }\textbf {\bibinfo {volume} {102}},\ \bibinfo
  {pages} {012616} (\bibinfo {year} {2020})}\BibitemShut {NoStop}%
\bibitem [{\citenamefont {Dodonov}\ \emph {et~al.}(1994)\citenamefont
  {Dodonov}, \citenamefont {Man'ko},\ and\ \citenamefont
  {Man'ko}}]{dodonov1994multi}%
  \BibitemOpen
  \bibfield  {author} {\bibinfo {author} {\bibfnamefont {V.~V.}\ \bibnamefont
  {Dodonov}}, \bibinfo {author} {\bibfnamefont {O.~V.}\ \bibnamefont
  {Man'ko}},\ and\ \bibinfo {author} {\bibfnamefont {V.~I.}\ \bibnamefont
  {Man'ko}},\ }\bibfield  {title} {\bibinfo {title} {Multidimensional hermite
  polynomials and photon distribution for polymode mixed light},\ }\href
  {https://doi.org/10.1103/PhysRevA.50.813} {\bibfield  {journal} {\bibinfo
  {journal} {Phys. Rev. A}\ }\textbf {\bibinfo {volume} {50}},\ \bibinfo
  {pages} {813} (\bibinfo {year} {1994})}\BibitemShut {NoStop}%
\bibitem [{\citenamefont {Avagyan}\ \emph {et~al.}(2022)\citenamefont
  {Avagyan}, \citenamefont {Knill},\ and\ \citenamefont
  {Glancy}}]{qc:avagyan2022}%
  \BibitemOpen
  \bibfield  {author} {\bibinfo {author} {\bibfnamefont {A.}~\bibnamefont
  {Avagyan}}, \bibinfo {author} {\bibfnamefont {E.}~\bibnamefont {Knill}},\
  and\ \bibinfo {author} {\bibfnamefont {S.}~\bibnamefont {Glancy}},\
  }\bibfield  {title} {\bibinfo {title} {Multi-mode {Gaussian} state analysis
  with total-photon counting}} (\bibinfo {year} {2022}),\ \bibinfo {note}
  {arXiv preprint arXiv:2209.14453}\BibitemShut {NoStop}%
\bibitem [{\citenamefont {Leonhardt}(1997)}]{leonhardt:qc1997a}%
  \BibitemOpen
  \bibfield  {author} {\bibinfo {author} {\bibfnamefont {U.}~\bibnamefont
  {Leonhardt}},\ }\href@noop {} {\emph {\bibinfo {title} {Measuring the Quantum
  State of Light}}}\ (\bibinfo  {publisher} {Cambridge University Press},\
  \bibinfo {address} {Cambridge, UK},\ \bibinfo {year} {1997})\BibitemShut
  {NoStop}%
\bibitem [{\citenamefont {Smith}(1995)}]{smith1995recursive}%
  \BibitemOpen
  \bibfield  {author} {\bibinfo {author} {\bibfnamefont {P.~J.}\ \bibnamefont
  {Smith}},\ }\bibfield  {title} {\bibinfo {title} {A recursive formulation of
  the old problem of obtaining moments from cumulants and vice versa},\ }\href
  {http://www.jstor.org/stable/2684642} {\bibfield  {journal} {\bibinfo
  {journal} {The American Statistician}\ }\textbf {\bibinfo {volume} {49}},\
  \bibinfo {pages} {217} (\bibinfo {year} {1995})}\BibitemShut {NoStop}%
\bibitem [{\citenamefont {Serafini}(2017)}]{serafini2017quantum}%
  \BibitemOpen
  \bibfield  {author} {\bibinfo {author} {\bibfnamefont {A.}~\bibnamefont
  {Serafini}},\ }\href {https://doi.org/10.1201/9781315118727} {\emph {\bibinfo
  {title} {Quantum Continuous Variables: A Primer of Theoretical Methods}}}\
  (\bibinfo  {publisher} {CRC Press},\ \bibinfo {year} {2017})\BibitemShut
  {NoStop}%
\bibitem [{\citenamefont {Lorentz}(1992)}]{lorentz1992multi}%
  \BibitemOpen
  \bibfield  {author} {\bibinfo {author} {\bibfnamefont {R.~A.}\ \bibnamefont
  {Lorentz}},\ }\href {https://doi.org/https://doi.org/10.1007/BFb0088788}
  {\emph {\bibinfo {title} {Multivariate {B}irkhoff Interpolation}}},\ Lecture
  Notes in Mathematics\ (\bibinfo  {publisher} {Springer Berlin, Heidelberg},\
  \bibinfo {year} {1992})\BibitemShut {NoStop}%
\end{thebibliography}%

\clearpage  

\onecolumngrid

\appendix
\appendixpage

\section{Explicit expression for the Wigner function of the \(n\) photon projector}
\label{wigner}

We determine the Wigner function \(W_{n}(\vec{r})\) for the projector onto
the \(n\)-photon subspace of \(S\) modes as a function of \(r^{2}\).
For $S=1$, we denote the Wigner function of \(\ketbra{n}\) by $W_n^{(1)}(\vec{r})$, and its form is given in the main text (MT) by MT Eq.~\eqref{eq:nwigner}.
By linearity of the Wigner transform that takes operators to functions in phase-space, $W_n(\vec{r})$ can be written as a linear combination of the Wigner functions of projectors onto the photon numbers in each mode. In particular,
\begin{align}
W_n(\vec{r}) & = \sum_{n_1+\hdots+n_S=n} \prod_{k=1}^S W^{(1)}_{n_k}(\vec{x_k}).
\end{align}
This relationship between the $S$-mode and the one-mode projectors can be captured compactly by expressing the generationg function in \(S\) modes, $M(\vec{r};z)= \sum_n W_n(\vec{r}) z^n$, in terms of the one-mode generating functions $M^{(1)}(\vec{x_k};z)= \sum_n W^{(1)}_n(\vec{x_k}) z^n$ of  each of the \(S\) modes. In particular,
\begin{align}
  M(\vec{r};z) & = \sum_{n}z^{n}W_{n}(\vec{r})  \nonumber\\
  &=\sum_n z^n \sum_{n_1+\hdots+n_S=n} \prod_{k=1}^S W^{(1)}_{n_k}(\vec{x_k})  \nonumber \\
& =  \sum_n  \sum_{n_1+\hdots+n_S=n} \left( \prod_{k=1}^S W^{(1)}_{n_k}(\vec{x_k}) z^{n_k} \right) \nonumber \\
& =  \prod_{k=1}^S \sum_{n_k} W^{(1)}_{n_k}(\vec{x_k}) z^{n_k}  \nonumber \\
& = \prod_{k=1}^S M^{(1)}(\vec{x_k};z). \label{laguerre_gen}
\end{align}
Then, substituting the Laguerre polynomials according to MT
Eq.~\eqref{eq:nwigner} in $M^{(1)}(\vec{x_k};z) $ and matching to
the generating function of Laguerre polynomials (MT
Eq.~\eqref{eq:glaguerre}), we arrive at
Eq.~(MT \ref{ngenfn}) 
\begin{align}\label{eq: M(r,z)}
M(\vec{r};z) &= \frac{e^{-r^2}}{\pi^S}\frac{1}{(1+z)^S}e^{2 z r^2/(1+z)}.
\end{align}

We can write $M(\vec{r};z)$ as
\begin{align}\label{eq: M(r,z)_p}
M(\vec{r};z) &= \frac{e^{-r^2}}{\pi^S} \frac{1}{(1+z)^{S-1}} L( 2r^2,-z) .
\end{align}
The generating function $\frac{1}{(1+z)^{S-1}}$ expands as
\begin{align}
\frac{1}{(1+z)^{S-1}} & = (1+(-z)+(-z)^2+\hdots)^{S-1} \nonumber \\
& = \sum_{k} (-1)^{k} \binom{k+S-2}{S-2} z^k.
\end{align}
By expanding \(L(2r^{2},-z)\), multiplying by the expansion
  of \(\frac{1}{(1+z)^{S-1}}\), and combining the terms with the same power of
  \(z\), we obtain
\begin{align}
W_n(\vec{r}) &=\frac{e^{-r^2}}{\pi^S} \sum_{k=0}^{n} (-1)^{k} \binom{k+S-2}{S-2} (-1)^{n-k} L_{n-k}(2r^2) \nonumber \\
&= (-1)^n\frac{e^{-r^2}}{\pi^S}  \sum_{k=0}^{n}   \binom{n-k+S-2}{S-2}L_{k}(2r^2). 
\end{align}
We substitute the explicit form of the Laguerre polynomials $L_k(x) = \sum_{l=0}^k \binom{k}{l} \frac{(-1)^l}{l!}x^l$ to obtain
\begin{align}
W_n(\vec{r}) & =   \frac{(-1)^n}{\pi^S}e^{-r^2} \sum_{k=0}^n \binom{S+n-k-2}{S-2}  \sum_{l=0}^k \binom{k}{l} \frac{(-1)^l}{l!}(2r^2)^l \nonumber \\
&=  \frac{(-1)^n}{\pi^S}e^{-r^2}  \sum_{l=0}^n \frac{2^l (-1)^l}{l!} r^{2l} \left[   \sum_{k=l}^n \binom{S+n-k-2}{S-2}  \binom{k}{l}     \right] \nonumber \\
&= \frac{(-1)^n}{\pi^S}e^{-r^2} \sum_{l=0}^n 2^l (-1)^l\binom{n+S-1}{l+S-1} \frac{r^{2l}}{l!}.\label{wn_explicit}
\end{align}
Here we used the binomial identity $  \sum_{k=l}^n \binom{S+n-k-2}{S-2}  \binom{k}{l} =  \binom{n+S-1}{l+S-1} $ in the third line. The polynomial \(P_{n}(r^{2})\) introduced in the MT is  given by
\begin{align}
  P_n(r^2) & =   e^{r^{2}}W_{n}(\vec{r}) \nonumber\\
   &=\frac{(-1)^n}{\pi^S} \sum_{l=0}^n 2^l (-1)^l\binom{n+S-1}{l+S-1} \frac{r^{2l}}{l!}.\label{pn_explicit}
\end{align}

\section{Computing the cumulants from the photon-number probabilities}

The probability of $n$ photons is given by
$p_n = (2\pi)^S\int dr^{2S} W_n(\vec{r}) W(\vec{r}) $.  According
  to the MT, define the probability distribution
  \(W'(\vec{r}) = 2^{S}e^{-r^{2}}W(\vec{r})/p_{0}\) and denote the
  \(n\)'th moment of \(r^{2}\) with respect to \(W'(\vec{r})\) by
  \(\mu_{n}\).  Substituting the expression for $W_n(\vec{r})$, we obtain
\begin{align}\label{eq: p_n_1}
  p_n &  = (-1)^n 2^S \sum_{k=0}^n \frac{ 2^k (-1)^k}{ k!} \binom{n+S-1}{k+S-1}  \int dr^{2S} r^{2k}    e^{-r^2} W(\vec{r})\nonumber\\
      &=p_{0}(-1)^n \sum_{k=0}^n \frac{ 2^k (-1)^k}{ k!} \binom{n+S-1}{k+S-1}  \int dr^{2S} r^{2k}    W'(\vec{r})\nonumber\\
       &=p_{0}(-1)^n \sum_{k=0}^n \frac{ 2^k (-1)^k}{ k!} \binom{n+S-1}{k+S-1} \mu_{k}.
\end{align}
The vector of relative photon-number probabilities $\bm{p} = (p_1/p_{0},\hdots,p_n/p_{0})^{T}$ is related to the vector of moments $\bm{\mu}=(\mu_1,\hdots,\mu_n )^{T}$ by a multiplication of the latter with a lower triangular matrix \(M = (M_{kl})_{k=1,l=1}^{n,n}\), where, according to Eq.~\ref{eq: p_n_1}, for \(k\geq l\) we have
  \(M_{kl}=(-1)^{k+l}\frac{2^{l}}{l!}\binom{k+S-1}{l+S-1}\). Because the diagonal
  of \(M\) is positive, \(M\) is readily inverted to compute \(\bm{\mu} = M^{-1}\bm{p}\),
  which gives the moments as a linear combination of the relative photon-number
probabilities \(p_{1}/p_{0},\ldots, p_{n}/p_{0}\).

The \(n\)'th cumulant associated with the moments of \(r^{2}\)  according
  to \(W'\) is denoted
  by \(f_{n}\).
The relationship between the moments and the cumulants of a measure is well-known and can be expressed by the following recursion relation \cite{smith1995recursive}:
\begin{equation}\label{eq: cum_and_mom}
f_{n} = \frac{\mu_n}{\mu_0} - \sum_{k=1}^{n-1} \binom{n-1}{k-1} f_{k} \frac{\mu_{n-k}}{\mu_0}.
\end{equation}
Thus, $f_{n} $ is a readily computable function of the
moments $\mu_1,\hdots,\mu_{n}$. To compute the cumulants from the
photon-number probabilities, we substitute the previously obtained
expressions for the moments in terms of the relative photon-number
probabilities.

\section{Expressing the cumulants as polynomials of modified normal parameters}\label{cumulant_comp}

The modified normal parameters are the tuples \((m_k, \lambda_k', c_k')\)
defined in the MT in terms of the  normal parameters \((m_k, \lambda_k, c_k)\)
according to $\lambda_k' = (1+1/\lambda_k)^{-1} $ and $c_k' = (c_k/\lambda_k)^2$. To express the cumulants as polynomials of the modified normal parameters we follow the same steps in the MT but with more detail.
We start with the generating function \(F(z)=\int d^{2S}\vec{r} e^{r^2 z}  W'(\vec{r})\) for the moments of $r^2$  (MT Eq.~\eqref{eq:F(z)}). We perform an orthogonal change of coordinates
so that in these coordinates (a) $\Gamma=\mathrm{diag}(\gamma_1,\ldots, \gamma_{2S})$ with the $\gamma_k$ in decreasing order, and (b) for a block of
identical $\gamma_k$, the corresponding $d_k$ are zero except for the
lowest index in the block, where they are non-negative. Then,
\begin{align}
\frac{p_{0}}{2^{S}}F(z) & = \int d^{2S}\vec{r} e^{r^2 z} e^{-r^2} W(\vec {r}) \nonumber \\
& = \int d^{2S}\vec{r} e^{r^2 (z-1)} \frac{1}{\pi^S} \frac{1}{\sqrt{\det(\Gamma)}} e^{-(\vec{r}-\vec{d})^T \Gamma^{-1}  (\vec{r}-\vec{d})} \label{gaussianform} \\
& = \frac{1}{\pi^S} \frac{1}{\sqrt{  \prod_{k=1}^{2S} \gamma_k}} \int d^{2S}\vec{r}  e^{r^2 (z-1)}  e^{ -\sum_{k=1}^{2S} \frac{  (r_k-d_k)^2 }{\gamma_k} } \nonumber \\
& = \frac{1}{\pi^S} \frac{1}{\sqrt{  \prod_{k=1}^{2S} \gamma_k}}  \prod_{k=1}^{2S}  \int dr_k  e^{r_k^2 (z-1)}  e^{ - \frac{  (r_k-d_k)^2 }{\gamma_k} } \nonumber \\
& = \frac{1}{\pi^S}\prod_{k=1}^{2S}  \frac{1}{\sqrt{  \gamma_k}}   \int dr_k e^{ - (1-z+\gamma_k^{-1})r_k^2 + 2d_k\gamma_k^{-1} r_k-d_k^2\gamma_k^{-1} }\label{eq:smf(z)}.
\end{align}
The integrals converge when the real parts of the coefficients of the $r_k^2$ in the exponents of the integrands are negative. This happens for the values of $z$ satisfying $\Re{z}-1-\frac{1}{\gamma_1}<0$. Since $\gamma_1>0$, $F(z)$ is well-defined in a neighborhood of $z=0$, and evaluates to
\begin{align}
  F(z) & =\frac{2^{S}}{p_{0}} \frac{1}{\pi^S}\prod_{k=1}^{2S}  \frac{1}{\sqrt{  \gamma_k}} \sqrt{ \frac{\pi}{(1-z+\gamma_k^{-1}) } } e^{ d_k^2 (\gamma_k+\gamma_k^2(1-z) )^{-1} -d_k^2\gamma_k^{-1} } \nonumber \\
       &=\frac{2^{S}}{p_{0}}\prod_{k=1}^{2S} \frac{ e^{-d_k^2\gamma_k^{-1}} }{\sqrt{(\gamma_k-z\gamma_k+1)}} e^{ d_k^2 \gamma_k^{-1} (\gamma_k-z\gamma_k+1)^{-1}  } 
\end{align}
We can thus evaluate the logarithmic derivative
$l(z) = (dF(z)/dz)/F(z)$ in such a neighborhood. We obtain
\begin{align}
l(z) & = \sum_{k=1}^{2S} \left[  -\frac{1}{2} \frac{1}{ (\gamma_k-z\gamma_k+1)}(-\gamma_k)-d_k^2 \gamma_k^{-1} \frac{1}{ (\gamma_k-z\gamma_k+1)^2} (-\gamma_k) \right] \nonumber \\
&= \sum_{k=1}^{2S} \left[  \frac{\gamma_k}{2} \frac{1}{ (\gamma_k-z\gamma_k+1)}+\frac{d_k^2 }{ (\gamma_k-z\gamma_k+1)^2}  \right].
\end{align}
In terms of the modified normal parameters $\lambda'_k = (1+1/\lambda_k)^{-1}$ and $c'_k = (c_k/\lambda_k)^2$ and the normal parameters $m_k$, the expression above becomes MT Eq.~\eqref{eq:lexpansion}:
\begin{align}
  l(z) &=
  \sum_{k=1}^h\left( \frac{\lambda'_k m_k}{2(1 - \lambda'_kz)} + \frac{\lambda'_k{}^2c'_k}{(1 - \lambda'_kz)^2}\right).
  \label{eq:lexpansion_sm}
\end{align}
We define \(f_{n}\) as the degree \(n-1\) coefficient of \(l(z)\), which
  relates to the \(n\)'th cumulant \(c_{n}\) of the distribution of \(r^{2}\) with respect to
  \(W'(\vec{r})\)  as \(c_{n}=(n-1)!f_{n}\).
The \(f_{n}\) can be determined from the power series expansions of the $(1 - \lambda'_iz)^{-1}$ and the $(1 - \lambda'_iz)^{-2}$, and collecting terms with the same powers of $z$. In particular
\begin{align}\label{eq: f_n}
  f_n &=\sum_{k=1}^s \left(\frac{m_k}{2} \lambda'_k{}^{n} + n c'_k   \lambda'_k{}^{n+1}\right),
\end{align}
Which is MT Eq.~\eqref{fn}.

\section{ Determining the normal parameters from the first $8S$ cumulants}
\label{determining}

Here we describe how to explicitly invert the polynomial
  expressions of the modified normal parameters in terms of the
  cumulants. A proof that the normal parameters are determined by the
first $8S$ cumulants is given in
the MT. The proof constructs the polynomial
$f(z) = \sum_{k=0}^{8S-1} f_{k+1} z^k$, and shows that if
a polynomial $q(z)$ has a degree that is at most $4S$ and satisfies
the property that the coefficients of $z^{4S},\hdots, z^{8S-1}$ in $q(z) f(z)$ are zero, then
$q(z) f(z)=q(z) l(z)$ has degree at most $4S-1$, and
$q(z)$ is divisible by the minimal degree polynomial
$q_0(z)=\prod_k(1-\lambda'_k z)\prod_{k:c'_k \ne 0}(1-\lambda'_kz)$.
The first task of the procedure for determining the normal
parameters is to find $q_0(z)$.
This task is accomplished by finding the minimal degree polynomial $q(z)$ with a constant term of $1$, such that $q(z) f(z)$ has degree at most $4S-1$. Then, $q_0(z) = q(z)$. We consider candidate polynomials $q_k(z) = 1+ \sum_{l=1}^k g_l z^l$ for each degree $k=1,\hdots, 4S$. Define \(g_{0}=1\). Then,
\begin{align} 
q_k(z) f(z) & = \sum_{l=0}^k g_l z^l \sum_{m=0}^{8S-1} f_{m+1} z^m \nonumber \\
& = \sum_{l=0}^k  \sum_{m=0}^{8S-1} g_l  f_{m+1} z^{l+m} \nonumber \\
& = \sum_{j=0}^{k+8S-1} \left[ \sum_{l=\max (0, j-8S+1 )}^{\min (j,k )} g_l f_{j-l+1}   \right] z^j,
\end{align}
where $g_0=1$. Starting with $k=1$, for each consecutive value of $k$ we check if a set of coefficients $g_1,\hdots,g_k$ exist that are the solution to the corresponding system of $4S$ linear equations:
\begin{align}
f_{j+1}+\sum_{l=1}^{k } g_l f_{j-l+1} = 0,\textrm{ for }j=4S,\hdots,8S-1.
\end{align}
This is equivalent to checking if the following matrix $A_k$ is full-rank:
\begin{align}
A_k = \begin{pmatrix}   
f_{4S+1} & f_{4S} &\hdots & f_{4S+1-k} \\
f_{4S+2} & f_{4S+1} &\hdots & f_{4S+2-k} \\
\vdots & \vdots & \ddots & \vdots \\
f_{8S} & f_{8S-1} &\hdots & f_{8S-k} 
\end{pmatrix}
\end{align}
We terminate the algorithm for the smallest value of $k $ for which the corresponding $A_k$ is not full-rank. Then, the coefficients $g_1,\hdots, g_k$ of $q_0(z)$ are obtained by picking the vector from the one-dimensional right-kernel of $A_k$ with a first entry of $1$.

The normal parameters are obtained from $q_0(z)$ as follows. Looking at the form of $q_0(z)$, one can see that the $\lambda'_k$ are given by the multiplicative inverses of the roots of $q_0(z)$, and $\lambda_k$ is determined by $\lambda'_k$. If the root corresponding to $\lambda'_k$ appears once, then the corresponding displacement $c_k=0$, and if it appears twice, then $c_k\neq 0$. The number of distinct roots is equal to $h$. Note that the $\lambda_k'$ are in a decreasing order by definition. Let us denote the number of roots of $q_0(z)$ appearing twice by $\bar{h}$. 
The parameters $c_k'$ and $m_k$ are obtained using the forms of the $f_n$ in Eq.~\ref{eq: f_n}. Let us define the vectors 
\begin{align}
\vec{z}_n = (\lambda_1'^{n},\hdots,\lambda_h'^{n}, n\lambda_{\sigma(1)}'^{n-1}, \hdots,n\lambda_{\sigma(\bar{h})}'^{n-1} )^T,\nonumber
\end{align}
where $\sigma(k)$ is the $k$'th smallest integer in $\{l: c_l \neq 0\}$. Let us also define the vector 
\begin{align}
\vec{\omega} &= (m_1/2,\hdots,m_h/2,c'_{\sigma(1)}\lambda_{\sigma(1)}'^2,\hdots,c'_{\sigma(\bar{h})}\lambda_{\sigma(\bar{h})}'^2 )  \nonumber \\
& = (m_1/2,\hdots,m_h/2,c^2_{\sigma(1)},\hdots,c^2_{\sigma(\bar{h})})  \nonumber 
\end{align}
Then, we can write $f_n$ as $f_n = \vec{z_n} \cdot \vec{\omega}$. Defining $\vec{\tilde f} = (f_1,\hdots, f_{h+\bar{h}})$, we have $\vec{\tilde f} = Z \cdot \vec{ \omega}$, where
\begin{align}
Z = \begin{pmatrix} \vec{z}_1^T \\ \vec{z}_2^T \\ \vdots \\ \vec{z}_{h+\bar{h}}^T \end{pmatrix}.
\end{align}
$Z$ is invertible, which is a standard result in Hermite interpolation theory, see, for example, Chap.~2 of \cite{lorentz1992multi}. Thus, one can obtain $\vec{ \omega}$ from $Z$ and from the first $h+\bar{h}$ cumulants. One immediately obtains the $m_i$ and the $c_i$ from the entries of $\vec{\omega}$.

\end{document}